# Multilayer Cryogenic Powder Filters with Low Parasitic Capacitance

Itishree Pradhan[1, *], Hao Li,[1] Alina Rupp[1,2], Yosuke Sato[1], Henri Vo Van Qui[1,3],
Miuko Tanaka[1], Toshiya Ideue[1], Erwann Bocquillon[2], and Masayuki Hashisaka[1, *]

[1] Institute for Solid State Physics, The University of Tokyo, 5-1-5 Kashiwanoha, Kashiwa 277-8581, Japan
[2] Physikalisches Institut, Universität zu Köln, Cologne, Germany
[3] LIMMS/CNRS-IIS, The University of Tokyo, 4-6-1 Komaba, Meguro-ku Tokyo, 153-8505, Japan
E-mail: itishree@issp.u-tokyo.ac.jp, hashisaka@issp.u-tokyo.ac.jp

We report the development of a cryogenic powder filter that simultaneously offers high attenuation of radio-frequency (RF) signals in the gigahertz (GHz) range and minimized parasitic capacitance to ground. Conventional powder filters, which consist of a signal line passing through a metal powder-filled housing, attenuate high-frequency signals via the skin effect. However, these designs often suffer from significant parasitic capacitance between the signal line and the grounded chassis, which can compromise the performance of sensitive measurement setups by limiting their frequency bandwidth. In this work, we demonstrate that a multilayer powder filter design effectively achieves both high RF attenuation and reduced parasitic capacitance. This solution suppresses sample heating due to the unintentional intrusion of RF signals through the wiring, without degrading the performance of the measurement setup.

The rapid progress of cryogenic quantum technologies [1-5] has driven the need for analog filters capable of strong RF attenuation at millikelvin temperatures. Traditional electronic low-pass filters [6,7] such as resistance-capacitance (RC) filters are often used to attenuate alternating current (AC) signals at frequencies below GHz. However, they typically do not exhibit ideal performance at and above GHz frequencies due to the influence of parasitic elements in actual devices, such as the parasitic capacitance of a resistor. Therefore, instead of lumped low-pass filters, experimentalists commonly use metal-powder to filter out RF microwave photons incoming from room temperature through the wiring to heat a cryogenic sample, thereby efficiently cooling down the sample [8-15].

Powder filters attenuate high-frequency signals due to the skin effect of the metal surface. Scattering powder around the signal lines increase the surface contributing to the skin effect and therefore enhances the attenuation. The structure is often constructed by filling the powder into a metal chassis and passing a central conductor through it [Figs. 1(a) and (b)]. While this setup is easy to assemble and offers high attenuation performance, it increases parasitic capacitance between the central conductor and the chassis as a side effect. In transport measurements, the parasitic capacitance restricts the design of the measurement system and, consequently, sometimes degrades the experimental capability. In one of the most typical cases, for example, the measurement bandwidth is limited by the RC circuit formed by the parasitic capacitance and the sample resistance. Powder filters with low parasitic capacitance would help mitigate this problem and be advantageous for enhancing cryogenic measurement technologies.

This paper reports a powder-filter design that achieves both significant attenuation of high-frequency signals and low parasitic capacitance. The filter features a multilayer structure, prepared by coating the central conductor and the inner side of the chassis with commercial metal-powder-containing epoxy (Eccosorb® CR-124, Laird Technologies) and forming a spatial gap between them [Fig. 1(c) and (d)]; Thus, we call it a 'layered filter.' The low dielectric constant of the gap area minimizes parasitic capacitance. Meanwhile, the epoxy coating enhances the skin effect, highly attenuating RF signals. Transmission spectroscopy using a vector network analyzer (VNA) shows a broadband attenuation of 60 dB for a 40 mm filter, for example. The parasitic capacitance of the layered filter is less than 10 pF, which is significantly lower than that of conventional filters, whose capacitance typically lies in the nF range. We demonstrate the advantage of the layered filter by examining it with a homemade transimpedance amplifier (TIA) [16-19]. The performance of the TIA was maintained at a high level due to the low parasitic capacitance, whereas the conventional filter significantly deteriorated the bandwidth. Thus, the multilayer structure of the powder filter solves problems induced by the parasitic capacitance in cryogenic quantum transport experiments.

Figure 1 summarizes our powder-filter structure and its assemblies in comparison with the conventional design using similar packaging parts. In this study, we fabricated both conventional and layered filters to compare their performance. The chassis equipped with SMA connectors is made from oxygen-free copper and has an internal space of 6 mm × 8 mm × $L$, where $L$ is the length of a filter. For the conventional one, we fully filled CR-124 in the chassis after wiring the central conductor of a copper wire (diameter: 0.25 mm) [Fig. 1(a) and (b)]. In contrast, for the layered filter, we filled CR-124 into a thin polymer tube (inner diameter: 2.2 mm, tube wall thickness: 0.25 mm) with a copper wire threaded through it; the wire diameter is 0.25 mm, the same as the conventional one. Then, we installed the tube into the chassis, whose inner

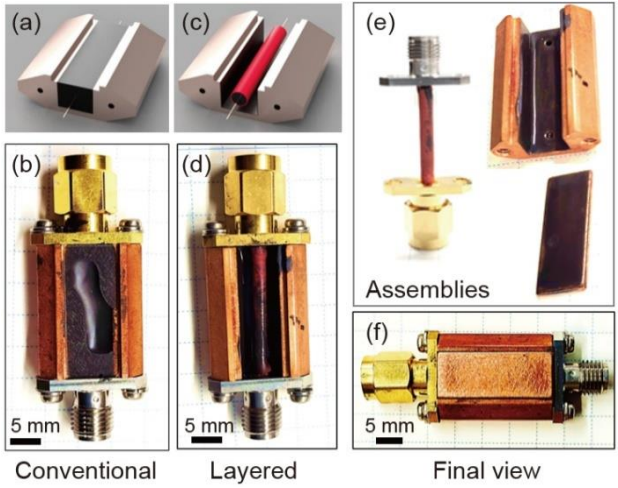

Fig.1 (a)(b) Schematics and photographs of the conventional powder filter of 20 mm wire length before covering. (c)(d) Those of the layered filter and (e) its assemblies. (f) Final view of the layered filter.

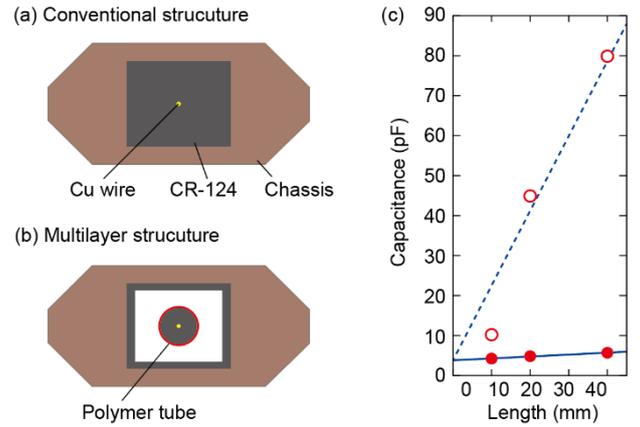

Fig. 2 (a)(b) Schematic cross-section of conventional and layered filters respectively. (c) Length dependence of parasitic capacitance. Open and filled circles are measurement data for the conventional and layered filters, respectively. Dashed and solid lines are linear fit for the data.

side is coated with CR-124 (typical thickness: 0.5 mm), by soldering the two ends of the wire to the SMA connectors [Fig. 1(c), (d), and (e)]. The final structure is obtained by setting a lid coated with CR-124 on the inner side [Fig. 1(f)].

The performance of these powder filters, such as the characteristic frequency and intensity of attenuation, depends on the particle size and density of the powder, as well as the length $L$. Therefore, we fabricated several conventional and layered filters of different lengths using different Eccosorb epoxies and compared their performances. In this paper, we report the $L$ dependence of the filters using CR-124 epoxy, while comparisons between different epoxies are available in the supplementary material.

While the final view of a layered filter [Fig. 1(f)] resembles a conventional one, it differs from the latter in the presence of a gap between the tube and the chassis inside. Figures 2(a) and (b) show the schematic cross-section of these filters. While the space between the central conductor and the chassis is filled with epoxy in the conventional case, the layered one has a space of typically 1.5 mm between the tube and the chassis. We measured the parasitic capacitance of the fabricated filters using a commercially available capacitance bridge (AH2700A, Andeen Hagerling). We swept the measurement frequency from 100 Hz to 1 kHz and found no frequency dependence in the results. Figure 2(c) displays the $L$ dependence of the parasitic capacitance of the conventional (open circles) and the layered filters (filled circles) measured at room temperature. While both filter structures show a monotonic increase in capacitance with $L$, the capacitance value (about 5 pF) of a layered structure is much smaller than that of a conventional one. The capacitance increases linearly with length $L$, with a slope of approximately 50 pF/m, which is comparable to the value calculated assuming the relative permittivity of air ($\varepsilon \sim 1$) for the 1.5 mm gap. Due to the small capacitance between the wire and the chassis, the total parasitic capacitance of the filter is primarily determined by that of the SMA connectors. On the other hand, the conventional structure has a slope of about 2 nF/m, about 40 times larger than the layered one due to the high dielectric constant of the epoxy and the presence of the metal powder.

The typical characteristic impedance of the layered filter is approximately 65 Ω, which is larger than 50 Ω, as expected from the lower capacitance between the central conductor and the chassis. The deviation from the 50 Ω characteristic impedance of standard coaxial cables causes the impedance mismatch in RF circuits [20-22]. On the other hand, as demonstrated later, this filter performs well without any restrictions in general cryogenic applications typically operating in the MHz band or below, where impedance matching is not strictly necessary.

We evaluated the attenuation characteristics of the filters using a VNA (N5224B, Keysight Technologies) with an input power of 0 dBm. Figure 3(a) shows the transmission spectra $S_{21}$ from 90 MHz to 43.5 GHz for both conventional and layered filters, each with a length of $L = 20$ mm, measured at room temperature. The layered filter exhibits significant broadband attenuation of 30 to 40 dB above 10 GHz, suitable for practical applications. In contrast, the conventional filter shows even stronger attenuation—about 80 dB or more—above 7 GHz, which is consistent with previous reports on similar filter structures [13-15]. The smaller attenuation of the layered filter compared to the conventional one can be understood due to the weaker skin effect resulting from the thinner powder layer.

Figure 3(b) shows the $S_{21}$ spectra of the layered filters of the different lengths. As expected, attenuation increases with the filter length, reaching approximately 60 dB at $L = 40$ mm. This characteristic is almost unchanged at 4.2 K [Fig. 3(c)], demonstrating the suitability of this filter for cryogenic use. As shown in Fig. 3(d), the attenuation due to the skin effect is negligible near 100 MHz. Thus, the layered filter of sufficient length has a high ability to attenuate microwave photons irradiated onto the sample through the wiring, while the skin effect does not affect the measurement quality at low frequencies.

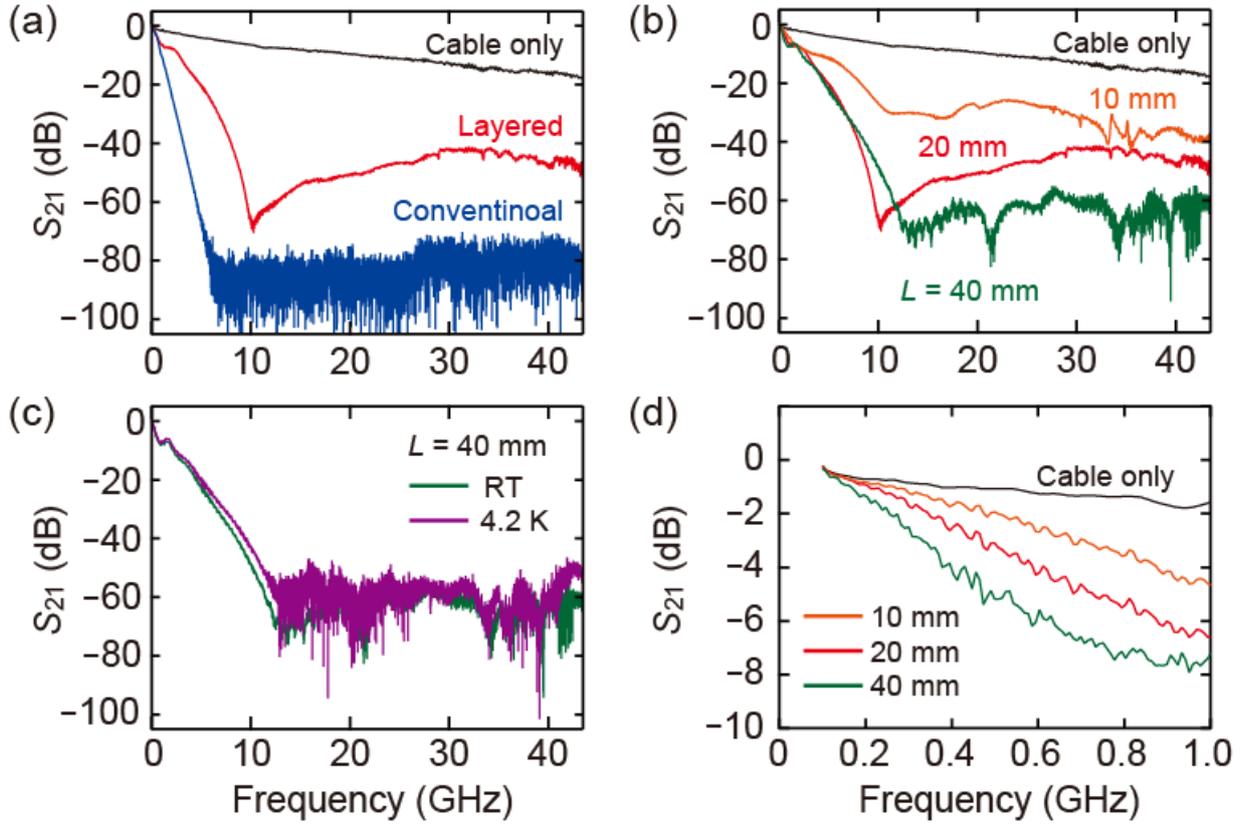

Fig. 3 Transmission spectra of fabricated powder filters. The black traces are the reference spectrum without a filter installed. (a) Comparison between layered (red) and conventional (blue) filters of $L$ = 20 mm. (b) Spectra of the layered filters of different lengths: 10 (orange), 20 (red), and 40 mm (green). (c) Comparison between the attenuation performance of 40 mm layered filters at room temperature and 4.2 K. (d) Low-frequency transmission spectra below 1 GHz at room temperature.

The advantage of the layered filter over the conventional one is its lower parasitic capacitance. We incorporated the layered filter into a cryogenic current measurement setup in the MHz band to examine its impact. Figure 4(a) displays the schematic diagram of the measurement setup. Assuming a high-precision measurement on a device under test (DUT) with a resistance of $R_{DUT}$ = 100 kΩ, we connected the output port of the DUT immersed in liquid helium to a homemade cryogenic TIA through a short SMA adapter and, in some cases, the powder filter [16-19]. The broadband and low-noise characteristics of the TIA are advantageous for high-precision measurement; the bandwidth and noise floor have a trade-off relationship, which depends on the feedback resistance $R_{FB}$ in the TIA circuit. In the present setup, we set $R_{FB}$ = 180 kΩ, where the upper limit of the flat band response is about 2 MHz, and the noise floor is about $2 \times 10^{-27}$ A$^2$/Hz. The details of the TIA design and its performance are available in Ref. [19].

We compared the performance of the setup [Fig. 4(a)] in the following three cases: (1) Direct connection between the DUT and the TIA, (2) the DUT-TIA connection with the conventional 20 mm filter inserted in between, and (3) the layered 20 mm filter inserted. We measured the bandwidth and the input-referred current noise $S_I$ of the TIA, which are standard indicators of the time resolution and accuracy, respectively.

Figures 4(b) and (c) respectively summarize the frequency dependence of the absolute value of the TIA transimpedance ($|Z_{trans}|$) and the phase shift of the TIA in the above three cases. Under the direct-connection condition, we observe $|Z_{trans}| \cong 180$ kΩ over the wide band, extending up to above 1 MHz, accompanied by a phase shift that remains near the ideal negative feedback value of −180 degrees. In this case, the total parasitic capacitance of the DUT and TIA input terminals and the adapter connecting them is approximately 10 pF.

The insertion of the conventional or layered filter does not vary the $|Z_{trans}|$ value in the flat band but reduces the upper limit of the bandwidth. The layered filter, with approximately 5 pF of parasitic capacitance, induces only a slight reduction, thereby maintaining system performance. This observation strongly contrasts with the conventional filter of about 50 pF, which significantly reduces the bandwidth by approximately an order of magnitude. These observations result from the fact that the parasitic capacitance at the connection between the DUT and TIA primarily determines the upper limit of the bandwidth. The difference between layered and conventional filters is also evident in the noise performance. In the cases of no filter or a layered filter inserted, the input-referred current noise is flat at approximately $2 \times 10^{-27}$ A$^2$/Hz up to 1 MHz, while it rises above this level near 300 kHz with the conventional filter [Fig. 4(d)]. Thus, we have demonstrated that a high-precision measurement is possible in a broader frequency band by adopting a layered filter instead of a conventional one. We expect that the small parasitic capacitance of the layered filter will contribute to improved measurement quality even in cases other than the setup shown in Fig. 4(a).

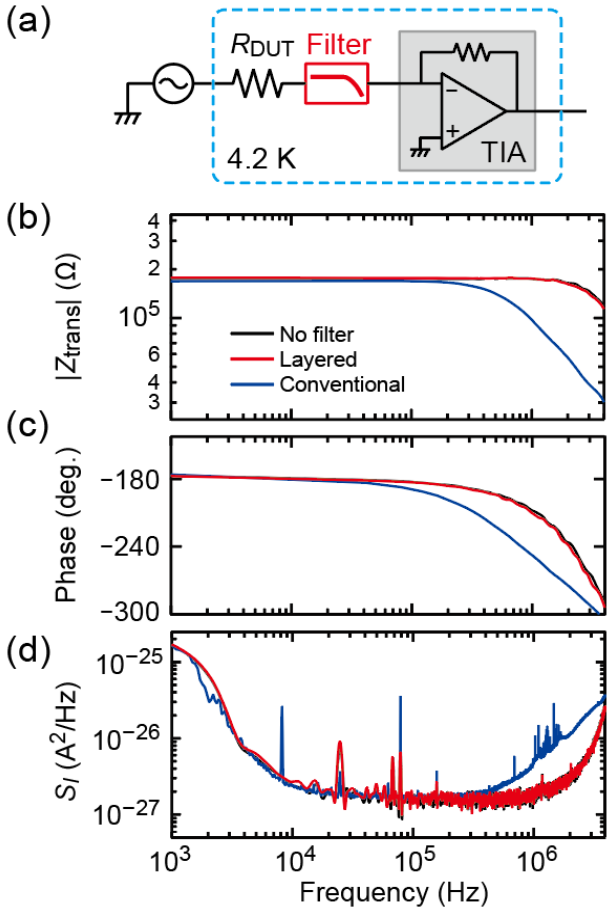

Fig. 4 Application example for a cryogenic transport measurement. (a) Schematic of the measurement setup. We examined the influence of fabricated filters inserted between DUT and TIA by evaluating the bandwidth and noise floor of the setup. (b) Bode plots of the TIA transimpedance $|Z_{trans}|$ and (c) the phase shift. (d) Spectral densities of the TIA input-referred current noise.

We have presented the design and performance of layered powder filters. Transmission spectroscopy has demonstrated a high attenuation performance (~ 40 dB) of the filters at GHz frequencies, indicating that they are promising for cryogenic applications to suppress microwave photons heating a sample. The advantage of the layered filter is its very low parasitic capacitance, approximately 40 times smaller than that of conventional one. It allows us to optimize measurement setups, particularly at low frequencies. For example, we have demonstrated broadband current measurement up to MHz frequencies with the layered filter installed. In conclusion, the multilayer filter design enables enhanced low-temperature measurement techniques and offers promising applications in emerging quantum technologies.

**Acknowledgements**

The authors appreciate the original design of the copper chassis by T. Fujisawa. We also thank M. Imai and H. Kamata for their technical support and fruitful discussions. This work was supported by Grants-in-Aid for Scientific Research (Grant Nos. JP22H00112, JP24H00827, and JP25H00613), JST ASPIRE program No. JPMJ1276308, the JSPS Bilateral Program No. JPJSBP120249911, UTEC-UTokyo FSI Research Grant Program, and Toray Science and Technology Grant.